\documentclass{article}
\usepackage{amsfonts}
\usepackage{amssymb}
\usepackage{epsfig}

\title{Feshbach Resonances and  Medium Effects  in ultracold atomic Gases \thanks{Presented at the 5th workshop on Critical Stability, Erice, Italy
13-17 October 2008}}
\author{G. M. Bruun\\The Niels Bohr Institute, \\University of Copenhagen,  DK-2100
Copenhagen \O, \\Denmark}

\begin{document}

\maketitle
\begin{abstract}
We develop an effective low energy theory for multi-channel scattering of cold atomic alkali atoms  with particular focus on Feshbach resonances.
 The scattering matrix is expressed in terms of observables only 
and the theory allows for the inclusion of many-body effects both in the open and in the closed channels.
 We then consider the frequency and damping of collective 
modes for Fermi gases and demonstrate how medium effects significantly increase  the scattering rate determining the nature of the modes. Our results 
obtained with no fitting parameters  are shown to compare well with experimental data.
\end{abstract}

\section{Introduction}
The study of cold atomic gases has now been at the forefront of low temperature physics for more than a decade. 
One can manipulate these gases with impressive experimental flexibility using the powerful tools of quantum optics. This has produced a string of ground breaking 
results relevant across many fields of physics including quantum optics, AMO and condensed matter physics~\cite{PethickBook,Giorgini}. 
A particularly attractive feature of cold atomic gases is the ability to manipulate the atom-atom interaction with the 
use of Feshbach resonances. The interaction can be made strong/weak and attractive/repulsive simply by tuning an external magnetic field. This  
has resulted in many important discoveries concerning strongly interacting many-body systems 
and the pace at which new results are being reported  shows no sign of slowing down.

 Sophisticated and very precise coupled  channels calculations have been developed to describe atomic Feshbach resonances at the two-body level~\cite{Kohler}. 
 Such coupled channels approaches are in general not easily generalized  to study  the intriguing many-body effects observed in the atomic gases. 
Several effective theories have therefore been developed which include a simplified version of the two-body Feshbach physics 
such that many-body calculations are tractable~\cite{Giorgini}. 
Most of these theories either neglect the Feshbach molecule entirely using so-called single channel 
models~\cite{Perali} or put it in  by hand as a point boson~\cite{Holland}. Such approaches 
have been very successful in calculating various many-body properties of the atomic gases for wide resonances 
where the multi-channel nature of the scattering is less important. For narrow resonances however, 
single channel approximations cannot be expected to be accurate, and even for wide resonances there are 
observables which depend specifically on the 
multi-channel nature of the scattering. 

To address this, we describe in this paper an effective theory for the Feschbach scattering  which  in the spirit of Landau expresses the multi-channel
 scattering matrix in terms of observables only. The Feshbach molecule  emerges dynamically as a 
 proper two-body state, yet the theory is still simple enough  to be easily generalized to treat many-body effects. 
As an application of this theory, we consider the collective modes of trapped atomic Fermi gases. 
The study of collective modes is a powerful probe into the properties of interacting quantum liquids. 
In cold atomic Fermi gases, collective mode spectroscopy has revealed a wealth of information about zero temperature $T=0$~\cite{Giorgini} as well as $T>0$ 
properties~\cite{Kinast,Wright,Riedl}. We outline how one can calculate the frequency and damping of the collective modes
 in the normal phase above the critical temperature $T_c$ for superfluidity. Focus is on how the modes 
 reveal information about the collisional properties and many-body effects.

\section{Landau Theory for in-medium  Scattering}
First we develop an effective low energy theory for fermionic alkali atom-atom scattering in a medium.  
 Consider alkali atoms in a magnetic field $B$ oriented along the $z$-direction. The strongest part of the atom-atom 
 interaction is the electrostatic central potential given by 
\begin{equation}
V(r)=\frac{V_s(r)+3\,V_t(r)}{4}+[V_t(r)-V_s(r)]\,{\vec{S}}_1\cdot{\vec{S}}_2
\label{vcentral}
\end{equation}
where $V_s(r)$ and $V_t(r)$ are the singlet and triplet potentials and ${\mathbf{S}}_1$ and ${\mathbf{S}}_2$ are the spins of the valence
electrons of the two alkali atoms~\cite{PethickBook}. Scattering via the potential (\ref{vcentral}) is characterized 
by channels of  anti-symmetrized two-particle states  with the same $z$-projection $M_z$ of the total spin $\vec{F}$.
 For a given $M_z$, the two-particle state with the lowest energy $\epsilon_{\alpha_2}+\epsilon_{\alpha_1}$ 
 constitutes the open channel $|o\rangle=|\alpha_1,\alpha_2\rangle$. Here 
 $\hat{H}_{\rm spin} |\alpha\rangle= \epsilon_\alpha|\alpha\rangle$ are the  eigenstates of the single particle hyperfine Hamiltonian~\cite{PethickBook}. 
The interaction (\ref{vcentral}) couples this channel to a number of higher energy states which 
form a set of closed channels $|c^{(n)}\rangle=|\alpha_3^{(n)},\alpha_4^{(n)}\rangle$.  
The threshold energies for the closed channels are then  
$E^{(n)}_{\rm th}(B) =\epsilon_{\alpha_4^{(n)}}+\epsilon_{\alpha_3^{(n)}} - \epsilon_{\alpha_2}-\epsilon_{\alpha_1}$ and they depend
on the magnetic field. 

\begin{figure}
\begin{center}
\epsfig{file=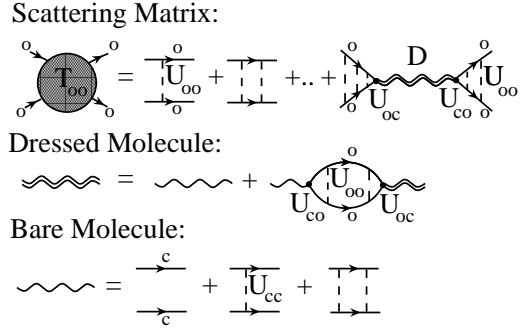, angle=0,width=0.6\textwidth}
\caption{The scattering matrix (\ref{Tatas}) decomposed into scattering in the open and  closed channels. The closed 
channel molecule is dressed via its coupling to the open channel. Fermions are indicated by straight lines, interaction within a channel is indicated by 
dashed lines, and coupling between the open and closed channels is indicated by $\bullet$. }
\label{Feynman}
\end{center}
\end{figure}

Focus now on the case where there is one open $|o\rangle$ and one closed 
scattering channel $|c\rangle$. It should be  emphasized that our effective theory is readily generalized to more than two channels if appropriate.  
To arrive at an effective theory for the scattering, we want to 
eliminate the bare microscopic interaction (\ref{vcentral})  which has a complicated momentum dependence. The high energy physics is eliminated 
by introducing an effective interaction $U_{ij}$ which is a solution to the zero energy Lippmann-Schwinger equation when the hyperfine
splitting of the channels is ignored. This results in a momentum independent low energy interaction given by~\cite{BruunKolomeitsev}
\begin{equation}
\hat{U}({\mathbf{q}}',{\mathbf{q}})=
\frac{4\pi}{m}\left[\frac{a_s+3a_t}{4}+(a_t-a_s){\mathbf{S}}_1\cdot{\mathbf{S}}_2
\right]
\label{EffectiveInteraction}
\end{equation}
where $a_s$ and $a_t$ are the scattering lengths for the singlet $V_s(r)$
and triplet $V_t(r)$ potentials, respectively. Any finite range  effects can be introduced through form factors which we have suppressed 
here for clarity. Using this low energy interaction, the Lippmann-Schwinger equation  
reduces to a simple $2\times2$ matrix equation
\begin{equation}
\left[
\begin{array}{cc}
T_{cc}&T_{co}\\
T_{oc}&T_{oo}
\end{array}
\right]^{-1}
=
\left[
\begin{array}{cc}
U_{cc}&U_{co}\\
U_{oc}&U_{oo}
\end{array}
\right]^{-1}
-
\left[
\begin{array}{cc}
\Pi_c&0\\
0&\Pi_o
\end{array}
\right]
\label{LippmannSchwinger}
\end{equation}
where $T_{ij}(\omega,{\vec{K}})$ is the scattering matrix between the channels $i$ and $j$. In addition to the usual dependence on the energy $\omega$,
   it also depends on the 
center-of-mass momentum ${\vec{K}}$ since  Galilean invariance is broken by the presence of the medium. 
The expressions for the pair propagators  in the open and closed channels
$\Pi_{0}(\omega,{\vec{K}})$ and $\Pi_{c}(\omega,{\vec{K}},B)$ with medium 
effects included through the ladder approximation
are given in Ref.~\cite{BruunKolomeitsev}. Equation (\ref{LippmannSchwinger}) is easily solved and the open channel scattering matrix 
can be written as
\begin{eqnarray}
T_{oo}=\frac{U_{oo}}{1-U_{oo}\Pi_o}+
\frac{U_{oc}}{1-U_{oo}\Pi_o}D\frac{U_{co}}{1-U_{oo}\Pi_o}
\label{Tatas}
\end{eqnarray}
 where 
 \begin{equation}
D^{-1}({\mathbf{K}},\omega)={\Pi_c}^{-1}-U_{cc}-{U_{oc}}^2\frac{\Pi_o}{1-U_{oo}\Pi_o}
\label{moleculeprop}
\end{equation}
is the in-medium pair propagator in the closed channel. 
Equation  (\ref{Tatas}) provides a transparent  physical interpretation of the multi channel scattering:
The first term in (\ref{Tatas}) describes scattering induced by the open channel interaction only and the second term
describes the scattering via the closed channel. 
The diagrammatic structure of (\ref{Tatas})-(\ref{moleculeprop}) is shown in Fig.\ \ref{Feynman}. 

The scattering of alkali atoms depends on the magnetic field $B$ 
both through the hyperfine states and the matrix elements $U_{ij}$.
 Close to a Feshbach resonance located at a given field $B_0$, the zero energy two-body scattering matrix
  in the open channel can be parametrized as 
\begin{equation}
T_{oo}^{vac}\equiv\frac{4\pi a}{m}=\frac{4\pi a_{\rm bg}}{m}\left(1-\frac{\Delta B}{B-B_0}\right).
\label{phenomenology}
\end{equation}
Here, $a_{\rm bg}$ is the (non-resonant) background scattering length and
$\Delta B$ the width of the resonance. 
The Feshbach resonance comes from the presence of a molecular state in the closed channel. It is thus contained in the second term in (\ref{Tatas}).
The energy $\omega_{\mathbf{K}}$ of a Feshbach molecule (including medium effects) with momentum ${\mathbf{K}}$ is
 determined by 
  $D^{-1}({\mathbf{K}},\omega_{\mathbf{K}})=0$. 
 By making a pole expansion of (\ref{Tatas}) around $B=B_0$ and comparing with (\ref{phenomenology}), one can write 
 the scattering matrix in the very useful form~\cite{BruunKolomeitsev}
 \begin{equation}
T_{oo}=\frac{T_{\rm bg}}{\left(1+\frac{\Delta\mu\Delta B}
{\tilde{\omega}+h(\omega)-\Delta\mu(B-B_0)}\right)^{-1}
-T_{\rm bg}\Pi_o(\omega)}.
\label{T(mu,abg)}
\end{equation}
Here $T_{\rm bg}=4\pi a_{bg}/m$, $\tilde{\omega}=\omega-K^2/4m$,
 and $\Delta\mu$ is the magnetic moment of the Feshbach molecule with respect 
to the open channel. A detailed analysis of the molecular propagator (\ref{moleculeprop}) shows that  $\Delta\mu$ can be split into a 
contribution from the magnetic dependence of the bare closed channel state
and a contribution from screening due to coupling to high energy states in the open channel. This screening which reduces
the magnetic moment from its bare value is often ignored in the literature. It  
comes from a \emph{linear} frequency dependence of the molecule self energy in addition to the well known
$\sqrt{\omega}$ threshold dependence, and it can lead to a significant reduction of the magnetic moment
 of the molecule~\cite{BruunKolomeitsev, bruunpethick}. The function $h(\omega)$ is given in Ref.~\cite{BruunKolomeitsev}. It describes effects coming from 
the composite two-fermion nature of the Feshbach molecule, and it is here that many-body effects in the closed channel enter.

With (\ref{T(mu,abg)}), we have arrived at an effective low energy theory for scattering in a medium.
 The complicated energy and momentum dependent  multichannel scattering matrix is  expressed in a simple way  through the physical 
observables $a_{\rm bg}$, $B_0$, $\Delta B$, and $\Delta\mu$. The parameters $a_{\rm bg}$, $B_0$, $\Delta B$ can be 
 measured in scattering experiments whereas the magnetic moment of the Feshbach molecule can be measured in rethermalization 
experiments ~\cite{BruunKolomeitsev}. Contrary to many other approaches in the literature, the theory 
allows one to include non-trivial many-body effects in the closed channel as well as in the open channel. 
Examples of such closed channel medium effects were considered in Ref.~\cite{BruunKolomeitsev}. 

\section{Collective Modes and Viscous Damping}
We now consider the collective modes of trapped Fermi gases and examine how they can reveal information about the scattering properties discussed in 
the previous section. We focus on the normal state for temperatures $T\ge T_c$.
The dynamics of the gas is assumed to be described by a semiclassical distribution function 
$f({\bf{r}},{\bf{p}},t)$ which satisfies the Boltzmann equation. A collective mode corresponds to a deviation 
$\delta f=f-f^0$ away from the equilibrium distribution $f^0({\mathbf{r}},{\mathbf{p}})$. By expanding
$\delta f$ in a set of basis states with the symmetry appropriate for the particular mode considered, one can express 
the Boltzmann equation in matrix form~\cite{Riedl,Massignan}. The corresponding determinants determine the 
mode frequency $\omega$. 

To be specific, we model the collective modes studied experimentally in Ref.~\cite{Riedl}, where the atoms are trapped in a very elongated harmonic
potential of the form $V({\bf r})=m(\omega_x^2x^2+\omega_y^2y^2+\omega_z^2z^2)/2$ with $\omega_z\ll \omega_y\le\omega_x$. The motion 
of the collective modes is then mainly in the $xy$-plane. For the scissors mode, the determinant equation determining the mode
frequency becomes ~\cite{Scissors}
 \begin{equation}
 \frac{i\omega}{\tau}(\omega^2-\omega_h^2)+(\omega^2-\omega_{c1}^2)(\omega^2-\omega_{c2}^2)=0.\label{Determinant}
 \end{equation}
Here $\omega_h=\sqrt{\omega_x^2+\omega_y^2}$  is the mode frequency in the hydrodynamic limit when $\omega\tau\ll 1$
characteristic of many collisions, and
$\omega_{c1}= \omega_x+\omega_y$ and $\omega_{c2}=|\omega_x-\omega_y|$ are 
the mode frequencies in the collisionless limit $\omega\tau\gg 1$~\cite{Odelin}. The collision rate $1/\tau$ is
\begin{equation}
\frac{1}{\tau}=\frac{\int d^3{r}d^3{p}p_xp_yI[p_xp_y]}{\int
d^3{r}d^3{p}p_x^2p_y^2f^0(1-f^0)}
\label{tau}
\end{equation}
where $I[p_xp_y]$ is the collision integral in the Boltzmann equation weighted by the momentum function 
$p_xp_y$~\cite{Riedl,Massignan}. It is in the collision integral, that the scattering matrix enters. 
The collision rate (\ref{tau}) is closely related to the viscosity of the gas and it is therefore sometimes called the 
viscous relaxation rate~\cite{Bruun}.  

\begin{figure}
\begin{center}
\leavevmode
\begin{minipage}{.49\columnwidth}
\includegraphics[clip=true,height=0.9\columnwidth,width=1\columnwidth]{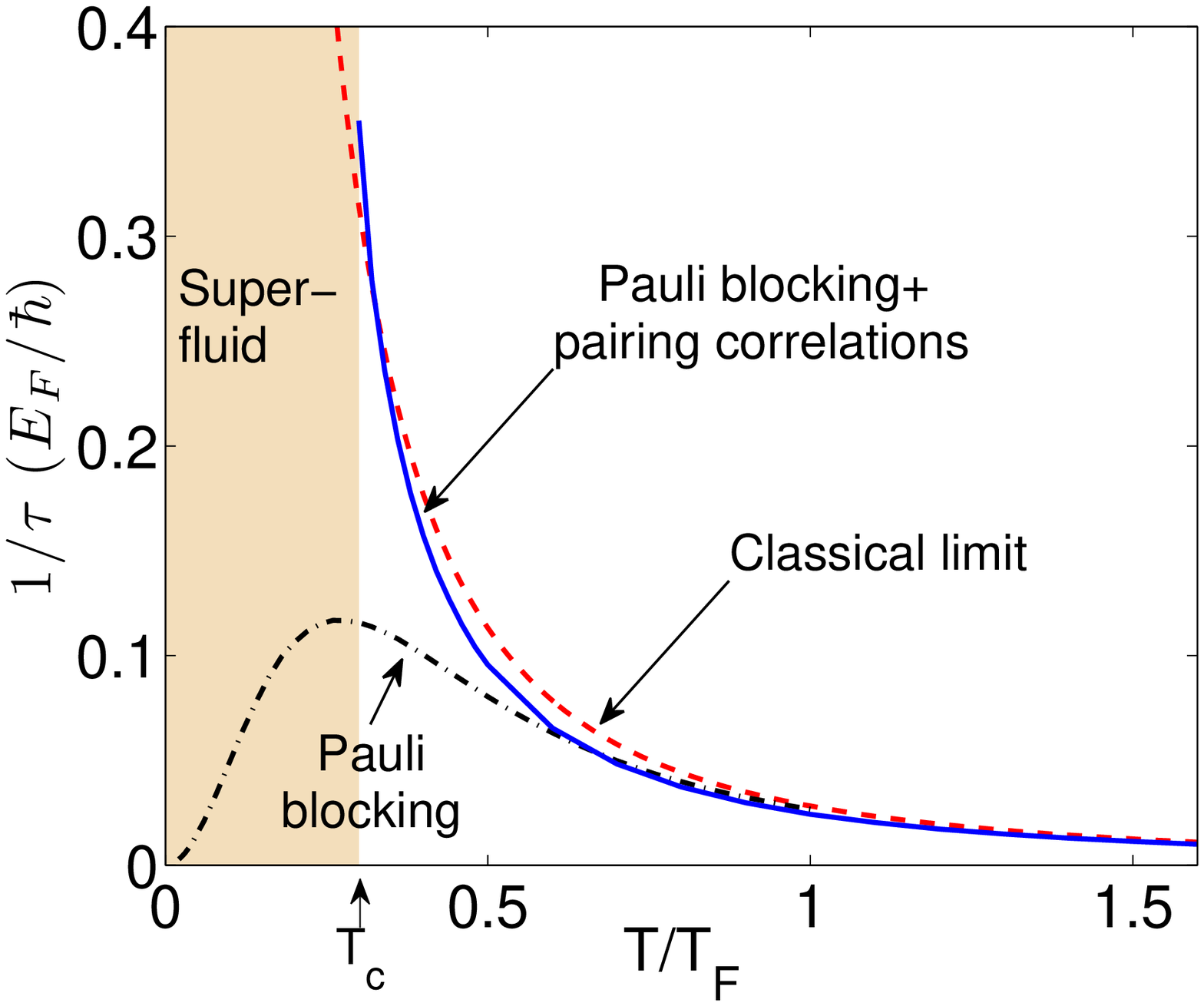}
\end{minipage}
\begin{minipage}{.49\columnwidth}
\includegraphics[clip=true,height=0.9\columnwidth,width=1\columnwidth]{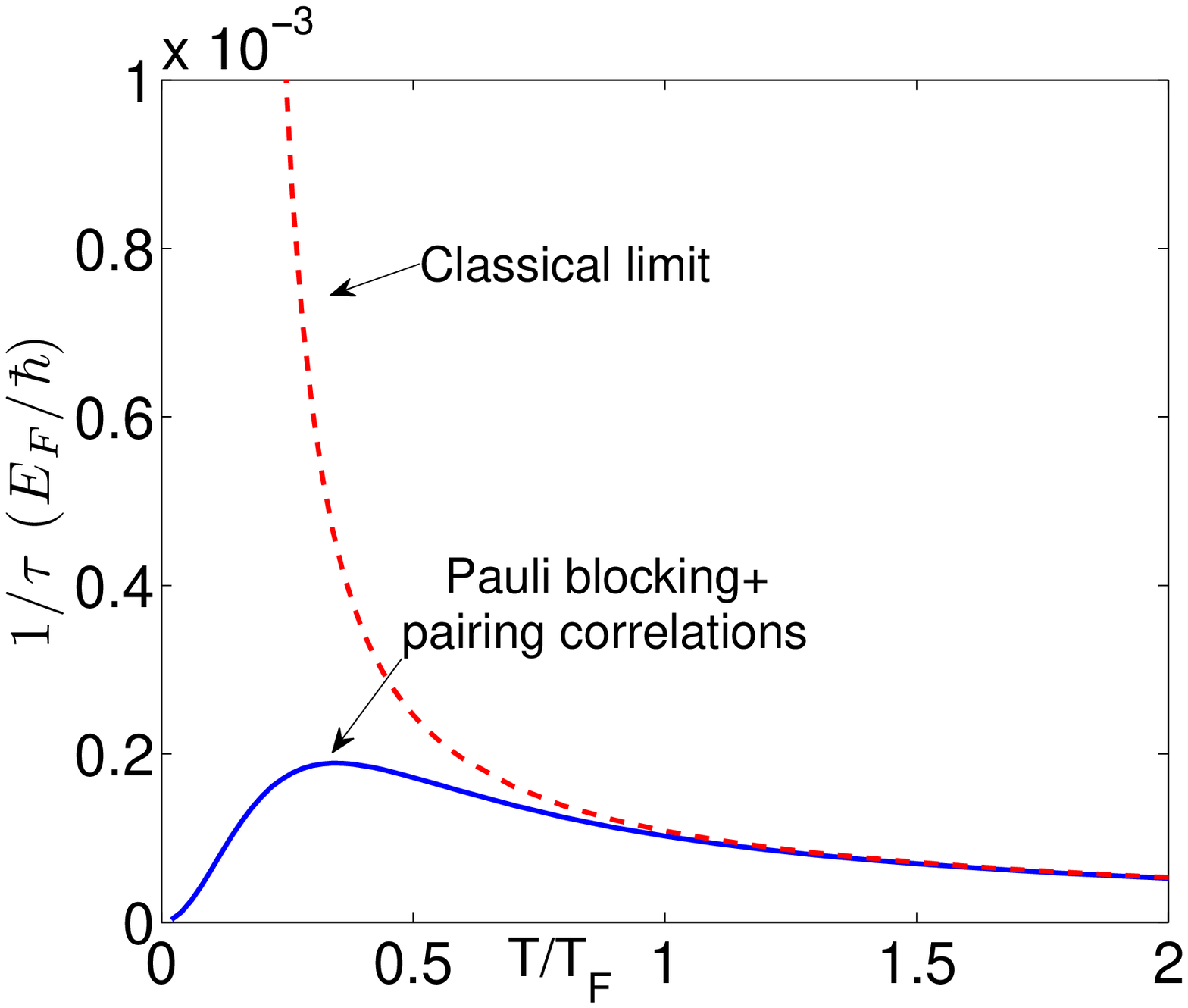}
\end{minipage}
\caption{The viscous relaxation rate rate $1/\tau$ for a gas in the unitarity limit (a) (From \cite{Riedl}) and in the 
weak coupling limit (b). The superfluid region is indicated in (a) whereas it is not visible in (b) due to the smallness of $T_c$. The dashed lines 
show the classical limit, the dash-dotted lines include Fermi blocking, and the solid lines include medium effects in the cross section.}
\label{Taufig}
\end{center}
\end{figure}

When the atoms are strongly interacting, there are 
significant pair correlations even in the normal phase. The correlations depend strongly on temperature
and  interaction strength which is parametrized by the scattering length $a$ in (\ref{phenomenology}). 
Correlations and their dependence on $a$ and $T$ enter the theory for collective 
modes through (\ref{tau}). In Fig.\ \ref{Taufig}, we plot the scattering rate $1/\tau$ as a function of $T$ for: (a) strong coupling right at 
a Feshbach resonance $|a|\rightarrow\infty$, and (b) in the weak coupling regime $k_Fa=-0.06$. 
To clearly identify  the importance of medium effects on the scattering matrix, 
the rate is calculated   using three different approximations. The dashed curves are a classical approximation where
Pauli blocking effects are neglected and the two-body scattering matrix is used. The dash-dotted lines include Pauli
blocking in the collision integral $I[p_xp_y]$ while the two-body scattering matrix is still used. We see that 
Pauli blocking reduces the scattering rate as compared to the classical result as expected; the classical rate scales as $T^{-2}$ whereas 
  $\tau^{-1}\propto T^2$ for $T\rightarrow0$ due to Pauli blocking. Finally, the solid lines use the many-body scattering matrix (\ref{T(mu,abg)}) 
  in addition to including Pauli blocking effects in the collision integral. Medium effects in the scattering matrix are 
 included through the  pair propagator $\Pi_{0}(\omega,{\vec{K}})$ in (\ref{T(mu,abg)}). As seen by comparing the solid and the dash-dotted lines in Fig.\ \ref{Taufig} (a), 
 medium effects significantly  increase the scattering rate over  a wide range of temperatures above $T_c$ for the strong coupling case. This is due to pair correlations. 
 It is the same physics which gives rise to a divergence in the ${\vec{K}}=0$ scattering matrix at $T_c$ signaling the onset of Cooper pairing. 
 From Fig.\ \ref{Taufig} (a), we see that the scattering rate calculated including both pair correlations and Pauli blocking effects is almost the same as 
 the classical rate which neglects both effects. Thus,  pair correlations nearly cancel the reduction of the scattering rate due to Pauli 
  blocking  in the normal phase. These strong pair correlations are often referred to as the pseudogap effect. 
   So we have demonstrated that it is essential to include medium effects in the scattering matrix  (\ref{T(mu,abg)}) when one 
  considers strong coupling Fermi gases; a simple two-body scattering matrix  strongly underestimates the correlations.  In contrast, 
  there are no observable medium effects on the scattering matrix  in the weak coupling regime depicted in Fig.\ \ref{Taufig} (b). Here the
  curves using a two-body and a many-body $T_{oo}$ are essentially indistinguishable.
  
\begin{figure}
\includegraphics[height=0.6\columnwidth,width=0.8\columnwidth]{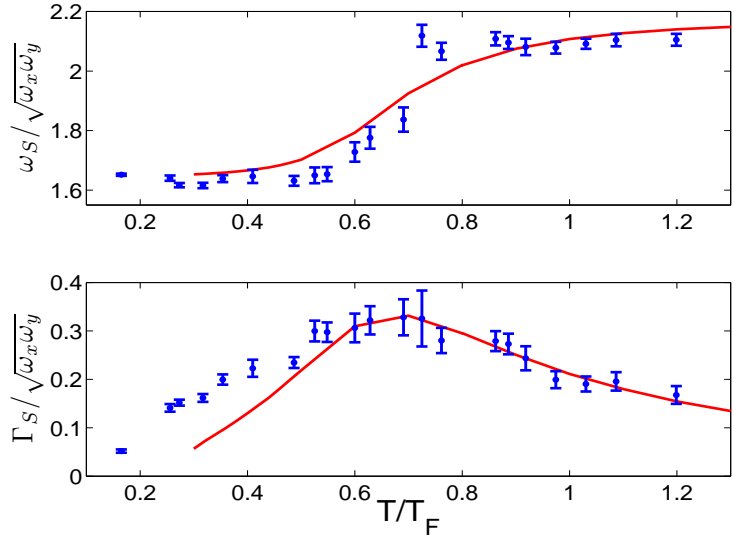}
\caption{The frequency $\omega_S$ and damping $\Gamma_S$ of the scissors mode. The solid line is theory and the experimental points are from 
Ref.~\cite{Riedl}.} 
\label{ScissorsMode}
\end{figure}

Once we know the scattering rate $1/\tau$, we can calculate the collective mode frequencies as discussed above. In Fig.\ \ref{ScissorsMode}, we
plot the scissors mode frequency $\omega_S$ and damping $\Gamma_S$ obtained from the real and imaginary parts of the solution of 
(\ref{Determinant}), i.e.\  $\omega=\omega_S-i\Gamma_S$. 
The scattering rate is obtained from (\ref{tau}) using the many-body scattering matrix (\ref{T(mu,abg)}).  The gas is strongly interacting with $|a|\rightarrow \infty$
and we compare with the experimental data in 
Ref.~\cite{Riedl}. Taking into account the experimental uncertainties and the fact that 
 there are \emph{no fitting parameters} 
in the theory, the agreement between theory and experiment is good. 
This indicates that the expressions 
 (\ref{T(mu,abg)}) and (\ref{tau}) account for most of the correlation effects  even for strongly correlated Fermi gases. 
Since the medium effects increase the scattering rate significantly, they  make the modes more hydrodynamic. We conclude that  
 the observation of well defined hydrodynamic modes above $T_c$ (See Fig.\ \ref{ScissorsMode} and Ref.~\cite{Riedl}) is a signature  of 
 many-body effects on the scattering. 
 
\section{Conclusion}
We developed an effective theory for multi-channel Feshbach scattering in cold alkali atom gases. The theory expresses the scattering in terms of 
physical observables only. It allows for the inclusion of many-body effects in all channels and provides a precise link
 between microscopic two-body multi-channel calculations and effective many-body theories. Many-body effects significantly increase the scattering rate over wide range of temperatures. We showed how this can be detected on the frequency and damping of collective modes. Our results were finally compared  to 
 experimental data obtaining good agreement.

\end{document}